\documentclass[aps,prl,reprint,superscriptaddress,twocolumn,floatfix]{revtex4-2}
\usepackage[T1]{fontenc}
\usepackage{textcomp}
\usepackage{amsmath}
\usepackage{amsfonts}
\usepackage{amssymb}
\usepackage{graphicx}
\usepackage{siunitx}
\usepackage{setspace}

\setcitestyle{number}

\usepackage[dvipsnames]{xcolor}
\newcommand{\rev}[1]{{\color{black} #1}}

\begin{document}

\title{Resonance-Induced Sign Reversal of Optical Gradient Forces
and Three-Dimensional Singularity Trapping}

\author{Jinsheng Lu}
\affiliation{%
 Harvard John A. Paulson School of Engineering and Applied Sciences,
 9 Oxford Street, Cambridge, Massachusetts 02138, USA
}%
\author{Soon Wei Daniel Lim}
\affiliation{%
 Harvard John A. Paulson School of Engineering and Applied Sciences,
 9 Oxford Street, Cambridge, Massachusetts 02138, USA
}%
\author{Federico Capasso}
\email{capasso@seas.harvard.edu}
\affiliation{%
 Harvard John A. Paulson School of Engineering and Applied Sciences,
 9 Oxford Street, Cambridge, Massachusetts 02138, USA
}%

\begin{abstract}
{In atomic physics, tuning the light frequency across a resonance reverses the trapping force between bright and dark field regions, yet a unified analytical description of this principle applicable to photonic resonators in general has not been established}.
Here we show that sweeping the incident wavelength through a resonance in the optical response of the particle or device induces a $\pi$ phase shift, {reversing the gradient force from attractive to repulsive.}{A generalized Fano line-shape model captures this quantitatively across physically distinct resonant systems, from plasmonic nanoparticles to high-Q metasurfaces}, with full-wave simulations confirming the predictions in every case. Building on this framework, three-dimensional singularity trapping of silicon nanoparticles is demonstrated using counter-propagating vector beams and metasurfaces, with trapping potential depths competitive with conventional bright-field traps. These results establish a platform-independent design principle for controlling optical forces in resonant systems, with broad implications for optical manipulation, quantum optomechanics, and precision metrology.
\end{abstract}

\maketitle

\textit{Introduction} --
Optical tweezers have revolutionized the manipulation of microscopic and nanoscopic objects, enabling significant advances in biology \cite{xu2024manipulating,ashkin1987optical,ashkin1970acceleration}, physics \cite{chu1991laser,chu1986experimental,Lukin_science,Lu_helicity_PRL,Lu_Polarization_PRL}, and chemistry \cite{heller2014optical,holland2023demand}. By exerting gradient forces through tightly focused laser beams, they have provided fundamental insights into molecular interactions \cite{ruttley_PRL}, cellular mechanics \cite{dao2003mechanics}, and quantum optomechanics \cite{rieser_science,hartung_Science,livska2024pt,kippenberg2008cavity}. Conventionally, particles are confined within high-intensity optical field regions by attractive (positive) gradient forces [Fig.~1(a)]. 

An alternative strategy involves trapping particles via repulsive (negative) optical gradient forces at dark regions or intensity minima [Fig.~1(b)], including optical singularities, which are field zeros \cite{lim2024multidimensional}. The topological and mechanical properties of optical singularities have motivated applications in photonic technologies \cite{spaegele2023topologically,shen2019optical,lim2021engineering, soskin2001singular,ni2021multidimensional,lim2024multidimensional}. Prior work has demonstrated repulsive gradient forces and dark-field trapping for low-refractive-index particles in engineered optical fields \cite{Melo_PRAppl_darkfocus,Almeida_PRL_darkfocus}, and for absorptive or high-index particles exploiting nonlinear effects \cite{qin2021nonlinearity} or resonance-induced negative polarizability, where the gradient force changes sign as the wavelength is swept across a resonance, reported predominantly for plasmonic particles \cite{ariasgonzalez2003optical,pelton2006optical,huang2008reversal,dienerowitz2008vortex,dienerowitz2008review,brzobohaty2015three,bendix2014optical,marago2013optical} and more recently for high-index dielectric Mie particles \cite{nietovesperinas2010optical,mao2025switchable,vernon2025nonlinear,Lepeshov_PRL_Metaatoms,afridi2025controllingsignopticalforces}. The closest analogy in physics is atomic trapping: tuning the laser frequency below (red-detuned) or above (blue-detuned) an atomic resonance switches the force between attractive and repulsive regimes, respectively \cite{davidson1995long,jarvis2018blue,Barredo_PRL}. Despite these advances, each study has addressed a specific platform or geometry in isolation, and the conditions governing gradient force sign reversal across engineered photonic systems of diverse physical origins have yet to be unified within a common analytical framework.

We show that this sign reversal is governed by a universal mechanism: sweeping the incident wavelength through an optical resonance induces a $\pi$-phase shift in the optical response of the particle or device, which reverses the gradient force and switches trapping between high- and low-intensity field regions. This phase-driven mechanism, analogous to the bright-to-dark transition in red- and blue-detuned atomic trapping, is independent of the physical platform and is captured, for the first time to our knowledge, by a single generalized analytical model based on Fano resonance line shapes that quantitatively describes force reversal and is confirmed by full-wave numerical simulations. We further demonstrate three-dimensional optical singularity trapping of silicon nanoparticles using counter-propagating radially polarized vector beams, and introduce wavelength-independent singularity trapping via metasurface-generated point-singularity arrays, overcoming the chromatic constraints of conventional counter-propagating geometries.

\begin{figure}[t]
\includegraphics{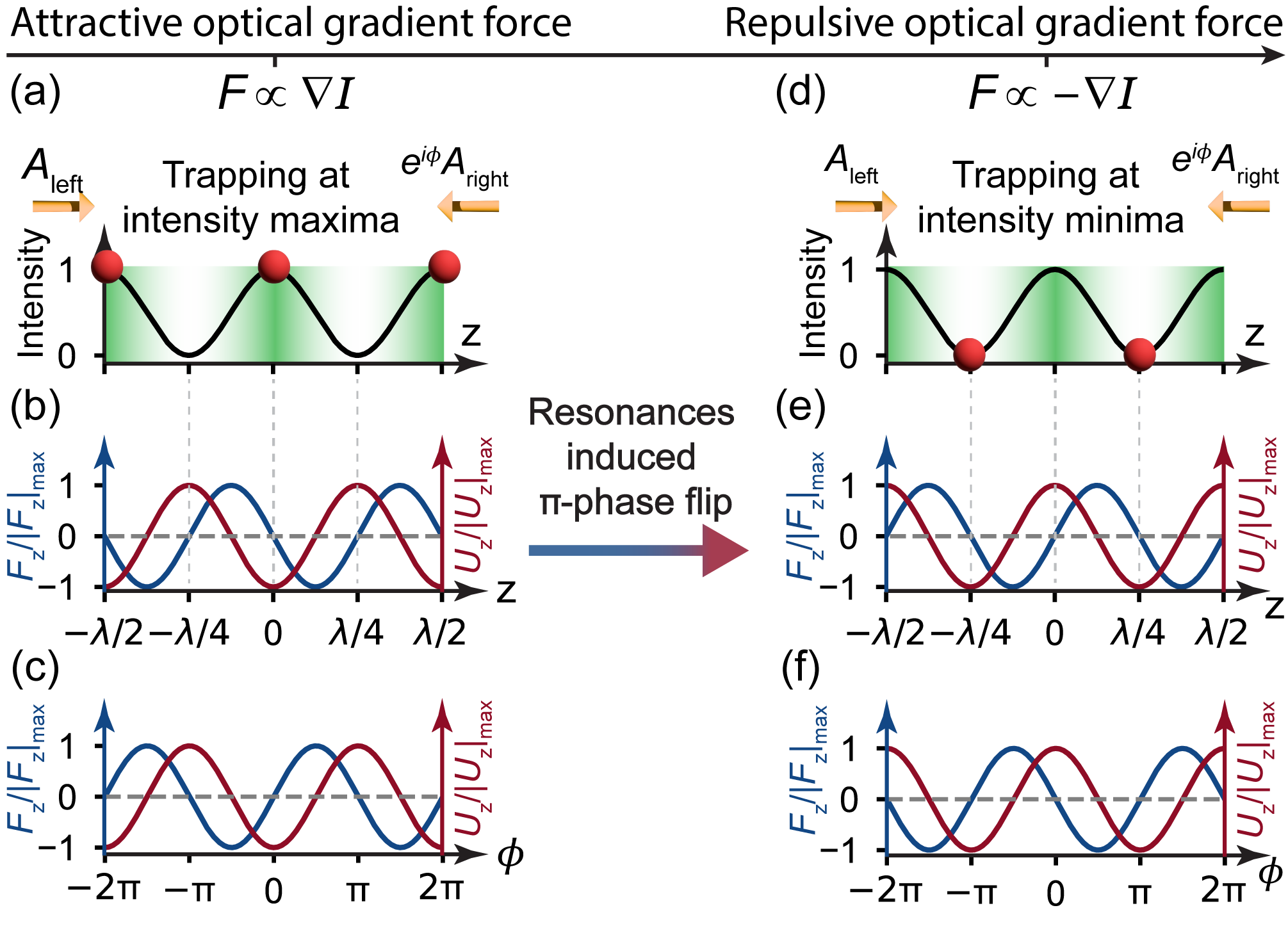}
\caption{\label{fig:fig1}
Principle of attractive and repulsive optical gradient forces in counter-propagating fields and their application to optical singularity trapping.} (a)--(c) Conventional attractive optical gradient force ($F \propto \nabla I$) and optical trapping at intensity maxima. (d)--(f) \rev{Repulsive optical gradient force} ($F \propto e^{i\pi}\nabla I = -\nabla I$) arising from a resonance-induced $\pi$-phase flip, leading to trapping at intensity minima/zeros (optical phase singularities). The counter-propagating beams have identical polarization and amplitude ($A_\text{left} = A_\text{right}$). (b),(c),(e),(f) Calculated optical force and trapping potential for $\phi = 0$ as the particle position is varied along $z$ [(b),(e)], and for a fixed particle position at $z = 0$ while sweeping $\phi$ [(c),(f)].
\end{figure}

\textit{Theoretical model} --
In the counter-propagating field configuration shown in Fig.~1(a), the intensity distribution is $I = \frac{1}{2}+\frac{1}{2}\cos\!\left(\frac{4\pi}{\lambda}z-\phi\right)$, where $\phi$ is the relative phase between the two beams and
$\lambda$ is the wavelength in the medium. The resulting gradient force on a test particle along the longitudinal $z$-direction is \cite{tlusty1998optical}:
\begin{equation}
    F_z \propto (\nabla I)_z =
    -\frac{2\pi}{\lambda}\sin\!\left(\frac{4\pi}{\lambda}z-\phi\right).
\end{equation}
Because the two beams propagate in opposite directions, net scattering forces cancel. The corresponding trapping potential is $U_z \propto -\tfrac{1}{2}\cos\!\left(\frac{4\pi}{\lambda}z-\phi\right)$. As shown in Figs.~1(b) and 1(c), particles experience an attractive gradient force and are confined at intensity maxima at $z_m = \!\left(\frac{m}{2}+\frac{\phi}{4\pi}\right)\!\lambda$, $m\in\mathbb{Z}$.

When the particle supports an optical resonance, its response acquires an additional spectral phase. For a single Lorentzian resonance, this phase is \cite{feynmanFeynmanLecturesPhysics2011a}: $\theta(\lambda) = \tan^{-1}\!\left(\frac{\lambda_r/\lambda}{Q(\lambda_r^2/\lambda^2-1)}\right)$, where $\lambda_r$ is the resonance wavelength and $Q$ is the quality factor. As $\lambda$ is tuned through $\lambda_r$, $\theta(\lambda)$ shifts by $\pi$, which inverts the gradient force:
\begin{equation}
    F_z \propto -(\nabla I)_z =
    \frac{2\pi}{\lambda}\sin\!\left(\frac{4\pi}{\lambda}z-\phi\right).
\end{equation}
Particles therefore experience a repulsive gradient force and are confined at intensity minima or zeros (optical phase singularities) [Figs.~1(d)--1(f)]. This behavior is unified across all resonant platforms by the following generalized force model:
\begin{align}
F_z(\lambda) &\propto
-\!\left(\sum_{n=1}^{N}\rho_n(\lambda)\right)
\frac{2\pi}{\lambda}\sin\!\left(\frac{4\pi}{\lambda}z-\phi\right),
\nonumber \\
\rho_n(\lambda) &= a_n + b_n
\frac{q_n+Q_n(\lambda_{n,r}/\lambda-1)}{1+jQ_n(\lambda_{n,r}/\lambda-1)},
\end{align}
where $\rho_n(\lambda)$ is a complex Fano resonance term that accommodates both symmetric Lorentzian and asymmetric spectral line shapes \cite{limonov2017fano}. The parameter $q_n$ governs the Fano asymmetry, while $a_n$ and $b_n$ are amplitude weights associated with the resonance wavelength $\lambda_{n,r}$. Eq.~(3) constitutes the unified model: the same functional form, with only platform-specific parameter values, describes gradient force reversal across all resonant systems in this work, from plasmonic nanoparticles to high-Q metasurfaces.

\begin{figure*}[t]
\includegraphics[width = 0.95\textwidth]{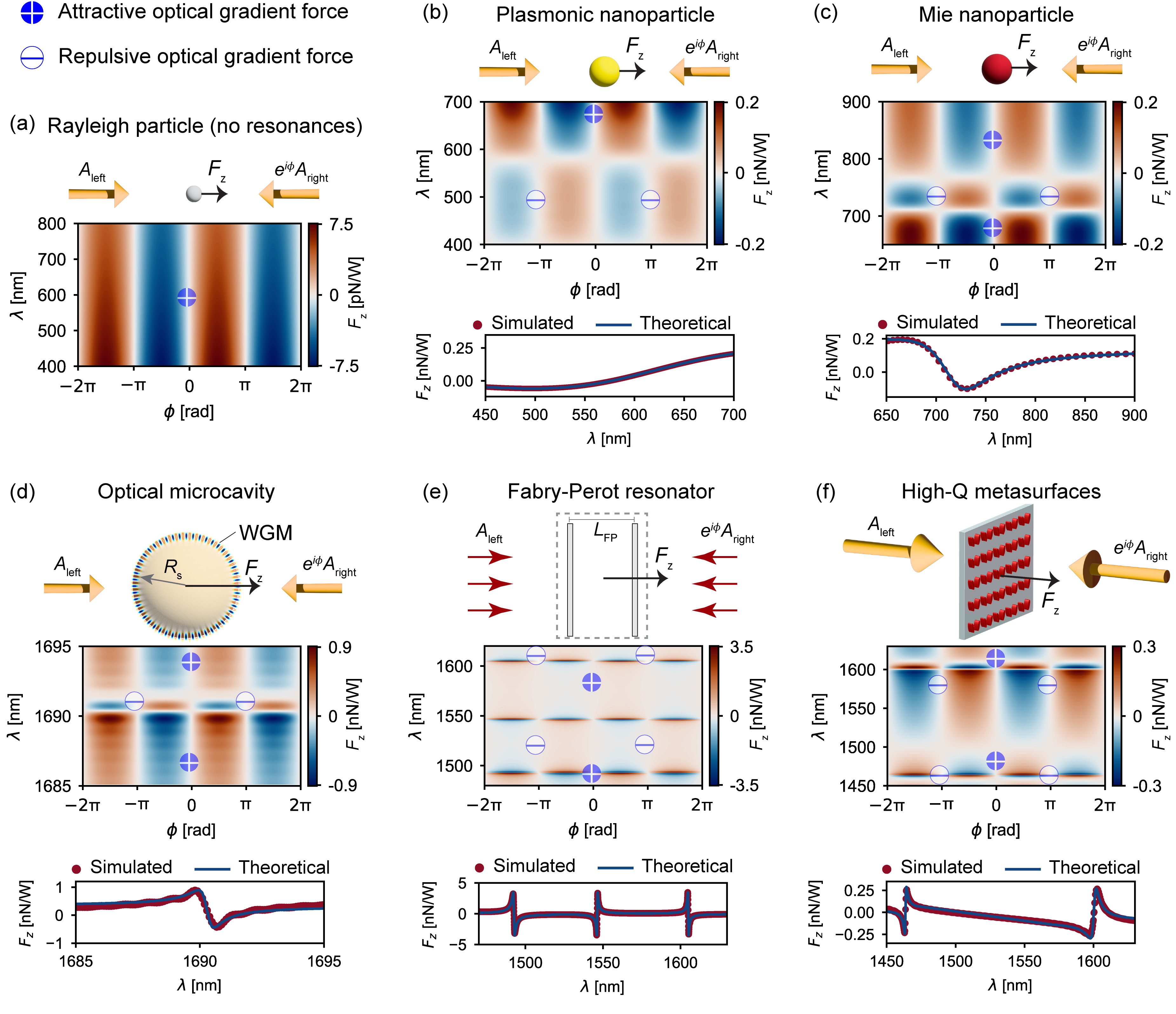}
\caption{\label{fig:fig2}
Demonstration of resonance-induced sign reversal of optical gradient forces across five distinct photonic platforms.} (a) Simulated total optical force $F_z$ on a silicon oxide nanoparticle (radius: 50~nm) in two counter-propagating beams as a function of phase difference $\phi$ and wavelength $\lambda$. (b)--(f) Corresponding force maps for (b) a gold nanoparticle (radius: 100~nm), (c) a silicon nanoparticle (radius: 100~nm), (d) a microcavity (radius: 3.3~$\mu$m, refractive index: 1.92), (e) a Fabry--P\'{e}rot resonator ($L_\text{FP} = 20~\mu$m, mirror reflectivity: 0.75), and (f) a silicon quasi-BIC metasurface. Curves below (b)--(f) are fits from Eq.~(3) at $\phi=\pi/2$. Parameter sets $[q, Q, \lambda_r, a, b]$:
$[-0.05, 4, 637~\text{nm}, 0.37, {-1.25}]$ for (b);
$[-1.1, 25, 718.6~\text{nm}, 0.81, 1.03]$ for (c);
$[-0.15, 5500, 1690.3~\text{nm}, 0.37, 1.5]$ for (d).
The Fabry--P\'{e}rot resonator in (e) has three resonances:
$[0, 3012, 1492.6~\text{nm}, 0, 2]$,
$[0, 6053, 1546.3~\text{nm}, 0, 2]$, and
$[0, 7136, 1605.0~\text{nm}, 0, 2]$.
The metasurface in (f) has two resonances:
$[0.04, 2590, 1463.7~\text{nm}, 0.13, {-2}]$ and
$[-0.35, 801, 1600.5~\text{nm}, 0.13, {-2}]$.
\end{figure*}

\textit{Unified framework across resonant systems} -- To validate the theoretical model, we perform three-dimensional finite-difference time-domain (FDTD) simulations for five representative resonant systems, spanning plasmonic, Mie-resonant, whispering-gallery, Fabry-P\'{e}rot, and quasi-BIC regimes, placed in counter-propagating fields. Time-averaged optical forces are computed via the Maxwell stress tensor (MST) and normalized to the total incident power.

As a non-resonant reference, a silicon dioxide nanoparticle in the Rayleigh regime is subject to an attractive gradient force across the entire wavelength range studied [Fig.~2(a)], consistent with Eq.~(1). In contrast, a gold nanoparticle (diameter: 200~nm) exhibits a plasmonic resonance that drives a $\pi$-phase flip in the gradient force as the wavelength crosses the resonance, producing a clear transition from attractive to repulsive force regimes [Fig.~2(b)]; repulsive trapping occurs when $[\lambda,\phi]=[400$ -$578~\text{nm},\,\pm\pi]$. A silicon nanoparticle supports multiple Mie resonances, yielding more than one $\pi$-phase flip in the studied wavelength range [Fig.~2(c)]. Extending the analysis to larger-scale photonic structures, we examine optical microcavities hosting whispering-gallery modes [Fig.~2(d)], Fabry-P\'{e}rot resonators [Fig.~2(e)], and high-Q metasurfaces supporting quasi-bound states in the continuum (quasi-BIC) and guided-mode resonances [Fig.~2(f)]. The resonant character of each platform is confirmed through its scattering cross section or transmission spectrum and the corresponding mode profile (Fig.~S1 and Sec.~S1 in the Supplemental Material). In systems where material absorption overlaps the resonant field distribution (such as particles coupled to whispering-gallery modes, whose physical extent spans many standing-wave periods), absorption-induced contributions should be assessed separately. Notably, in a standing wave the absorbed power need not follow the local intensity, so that a small particle trapped at an intensity minimum can nonetheless experience strong absorption, depending on whether the resonance is electric- or magnetic-dipole in character (Figs.~S9--S12 and Sec.~S4 in the Supplemental Material).

The analytical curves from Eq.~(3) are overlaid on the force maps in Figs.~2(b)--2(f) (for $\phi = \pi/2$), demonstrating
quantitative agreement with the FDTD results across all five platforms. The same functional form, with only platform-specific parameter values, reproduces the full force response in each case, distinguishing this unified treatment from prior platform-specific analyses \cite{Lepeshov_PRL_Metaatoms,afridi2025controllingsignopticalforces}. Good agreement also holds for the force spectra as functions of $\phi$ (at $z=0$, Fig.~S2) and $z$ (at $\phi=0$, Fig.~S3). These results establish that resonance-induced sign reversal of the optical gradient force is a universal phenomenon governed by the spectral phase of the resonant mode, independent of the physical platform.

\textit{Three-dimensional singularity trapping using vector beams} -- We next demonstrate three-dimensional optical singularity trapping as a direct application of this mechanism. All calculations in this section assume an aqueous environment. Because the counter-propagating beams carry equal and opposite linear momentum, their non-conservative (radiation-pressure) contributions largely cancel, leaving an essentially conservative force field. The trapping potential is therefore well-defined and is obtained by integrating the net optical force along each direction. Any residual non-conservative force is further suppressed by viscous damping, whereas vacuum trapping would require separate treatment of such contributions \cite{rieser_science}.

Two counter-propagating radially polarized vector beams are employed to generate isolated three-dimensional point singularities, in which the surrounding field intensity is non-zero while the central intensity vanishes. The beams show a nonparaxial V-type polarization singularity at their common focus \cite{lim2024multidimensional}; at this point both transverse and longitudinal polarization components simultaneously vanish, producing a characteristic donut-shaped intensity profile in the transverse plane [Fig.~3(a)]. In the longitudinal direction, standing wave formation and polarization structure give rise to a periodic array of dark points [Fig.~3(b); Fig.~S4 and Sec.~S2], whose spacing increases with wavelength [Fig.~3(c)]. The central singularity at $z = 0$ is wavelength-invariant, making it a robust trapping site across a broad spectral range.

To confirm three-dimensional trapping, we compute the optical forces and trapping potentials for a silicon nanoparticle (radius:
150~nm) scanned along the longitudinal axis at $(x,y)=(0,0)$ [Figs.~3(d) and 3(e)] and along the radial direction at $z=0$
[Figs.~3(f) and 3(g)]. In both cases, the resonance-induced $\pi$-phase flip is observed, switching the gradient force from attractive to repulsive. At $\lambda = 720$~nm, which coincides with the silicon Mie resonance, singularity trapping is achieved simultaneously in both directions [Figs.~3(h) and 3(i)]: the trapping potential minima align precisely with the intensity minima. The resulting potential depths are $\Delta U_z = 6.4~\mathrm{k_B T/mW}$ (longitudinal) and $\Delta U_r = 11.2~\mathrm{k_B T/mW}$ (radial), values comparable to or exceeding those reported for conventional bright-field trapping \cite{lehmuskero2015laser,Lepeshov_PRL_Metaatoms}.

\begin{figure}[t]
\includegraphics{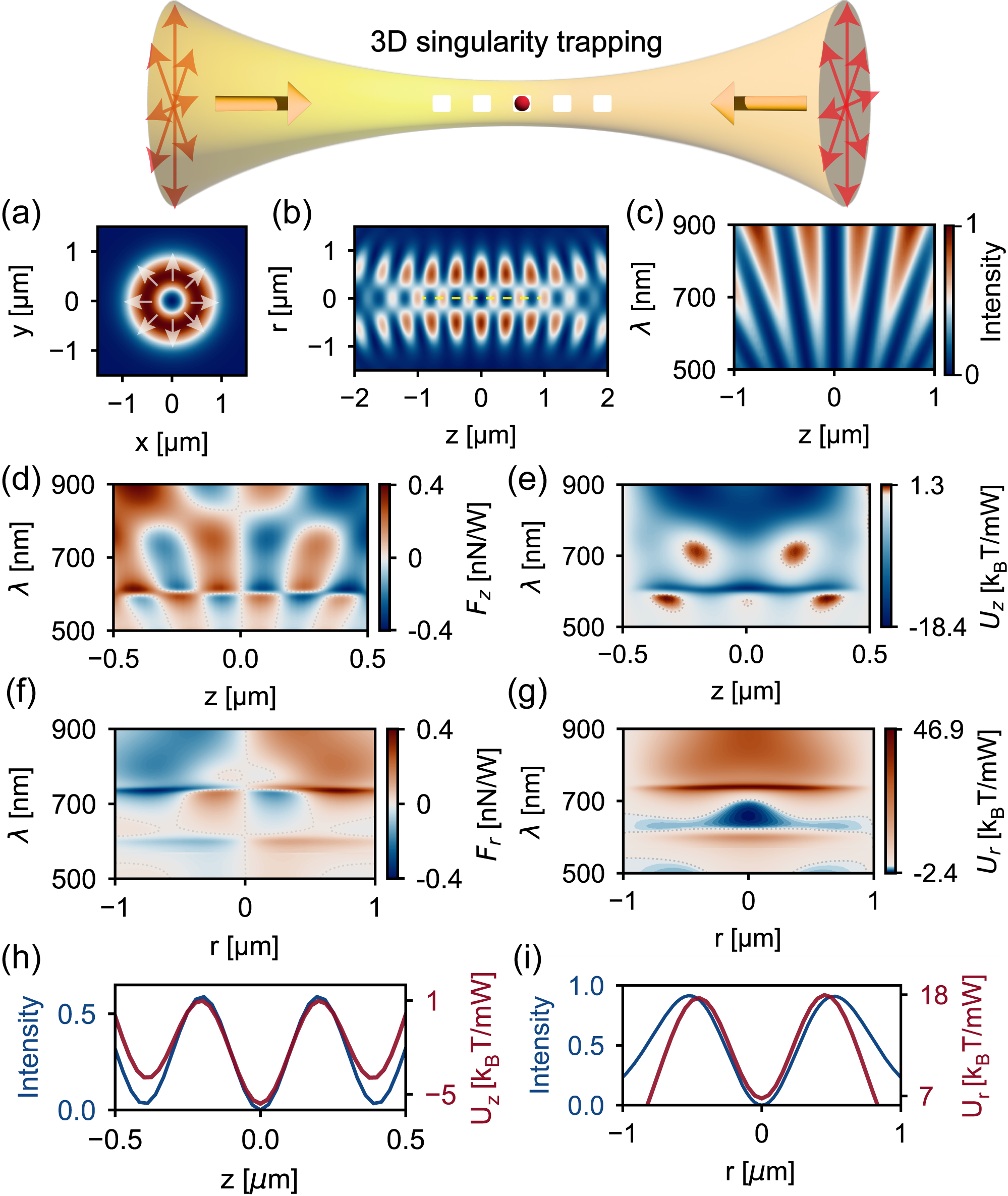}
\caption{\label{fig:fig3}
Three-dimensional singularity trapping of silicon nanoparticles using counter-propagating radially polarized vector beams.
(a)--(c) Intensity distribution in the transverse plane (a), longitudinal plane (b) at $\lambda = 720$~nm, and as a function of wavelength along the line indicated by the yellow dashed line in (b) (c). (d)--(g) Calculated optical force (d),(f) and trapping potential (e),(g) for a silicon nanoparticle (radius: 150~nm) along the longitudinal axis at $(x,y)=(0,0)$ [(d),(e)] and along the radial direction at $z=0$ [(f),(g)]. Red (blue) coloring denotes positive (negative) force; zero crossings are white. (h),(i) Intensity and trapping potential along the longitudinal (h) and radial (i) directions at $\lambda = 720$~nm.}
\end{figure}

\begin{figure}[t]
\includegraphics{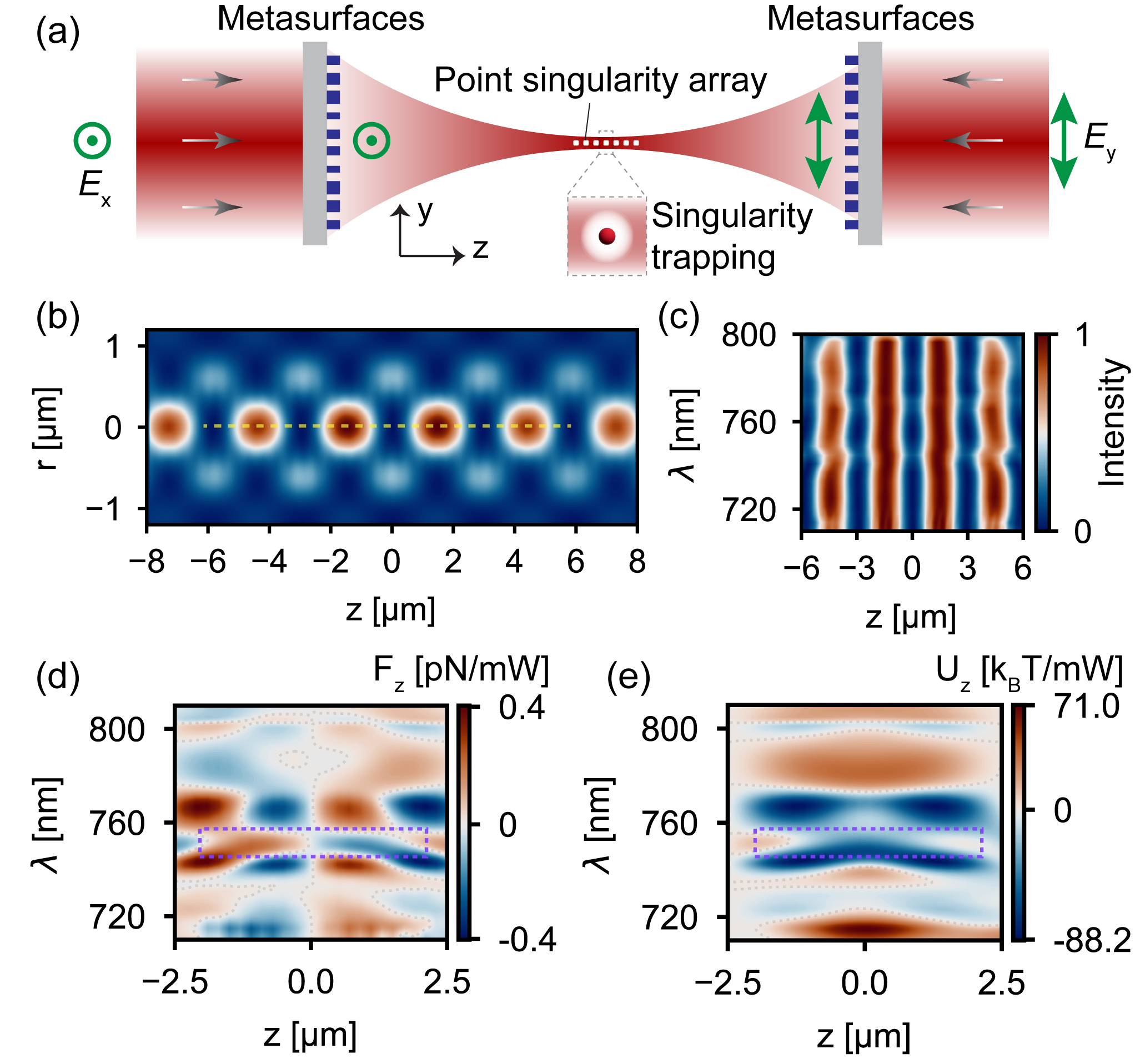}
\caption{\label{fig:fig4}
Optical singularity trapping of silicon nanoparticles using a metasurface-generated point-singularity array. (a) Schematic of the metasurface trapping configuration with orthogonally polarized counter-propagating beams. (b),(c) Intensity distribution of the metasurface-generated field in the longitudinal plane (b) at 757~nm, and as a function of wavelength along the line indicated by the yellow dashed line in (b) (c). (d),(e) Calculated gradient force $F_z$ (d) and trapping potential $U_z$ (e) for a silicon nanoparticle (radius: 175~nm). The resonance-induced $\pi$-phase flip near 750~nm produces singularity trapping (purple dashed rectangle).}
\end{figure}

\textit{Singularity trapping using achromatic metasurfaces} -- The singularities produced by counter-propagating vector beams suffer from a fundamental chromatic limitation: their longitudinal periodicity equals half the wavelength and their positions shift with wavelength, complicating broadband or wavelength-tunable operation. Additionally, their sub-wavelength dimensions restrict the size of particles that can be fully confined.

Metasurfaces can circumvent both limitations by generating a spatially fixed singularity array whose positions are insensitive to wavelength [Fig.~4(a)]; the metasurface design is detailed in Ref.~\cite{lim2023point} and Sec.~S3. The metasurface geometry enables control over the singularity size and periodicity, here producing singularities with characteristic dimensions approximately four times the wavelength [Fig.~4(b)]. As shown in Fig.~4(c), the singularity positions remain fixed as the wavelength is varied.

The counter-propagating beams are configured with orthogonal polarization states, suppressing standing-wave formation and strongly
reducing the net scattering force. The two metasurfaces are co-aligned along the optical axis to ensure precise spatial overlap of their respective singularity arrays, maximizing trapping stability. The calculated gradient force [Fig.~4(d)] and trapping potential [Fig.~4(e)] for a silicon nanoparticle (radius: 175~nm) display a resonance-induced $\pi$-phase flip in the range 748--755~nm, producing stable single-particle singularity trapping. Extensions to multi-particle trapping would require treatment of optical binding arising from inter-particle multiple scattering, and is left for future work.

\textit{Conclusion} -- We have presented a unified framework for optical gradient forces in resonant systems, establishing that the sign of the gradient force is governed by the spectral phase accumulated across a resonance. A single generalized analytical model based on Fano resonance line shapes [Eq.~(3)] quantitatively describes this force reversal across several physically distinct platforms: plasmonic nanoparticles, Mie-resonant nanoparticles, whispering-gallery microcavities, Fabry-P\'{e}rot resonators, and high-Q metasurfaces. This general principle provides a systematic design pathway for force-reversible optical trapping. We have demonstrated its application in three-dimensional singularity trapping of silicon nanoparticles, achieving trapping potential depths competitive with conventional bright-field traps. Resonant particles, including plasmonic metals and high-index semiconductors, for which conventional trapping is complicated by radiation pressure imbalances and optomechanical back-action, are natural candidates for dark trap manipulation via this mechanism. The framework also applies to macroscopic resonant systems such as optomechanical mirrors, with broad implications for precision force sensing, quantum optomechanics, and nanophotonic device integration.

\textit{Acknowledgment} --
We acknowledge support from the Air Force Office of Scientific Research under award numbers FA9550-21-1-0312 and FA9550-22-1-0243. S.W.D.L.\ is supported by the Schmidt Science Fellows, in partnership with the Rhodes Trust. S.W.D.L.\ is also supported by A*STAR Singapore through the National Science Scholarship scheme.

\bibliography{ost}

\end{document}